\documentclass[11pt,twocolumn]{article}

\usepackage{geometry}
\usepackage{cite}
\usepackage{amsmath,amssymb,amsfonts}
\usepackage{algorithm}
\usepackage{algorithmic}
\usepackage{graphicx}
\usepackage{textcomp}
\usepackage{xcolor}
\usepackage{booktabs}
\usepackage{hyperref}
\usepackage{url}
\usepackage{multirow}
\usepackage{array}
\usepackage{enumitem}
\usepackage{parskip}
\usepackage{caption}
\usepackage{float}

\hypersetup{colorlinks=true,linkcolor=blue,citecolor=blue,urlcolor=blue}
\begin{document}


\title{\textbf{LLM-CEG: Extending the Classification Error Gauge\\
Framework for Privacy Auditing of Large Language Models}}

\author{
  Kato Mivule \quad | Computer Science Dept. Bowie State University | \quad kmivule@gmail.com
}

\date{}

\maketitle


\begin{abstract}
This paper extends the Classification Error Gauge (x-CEG) framework,
originally developed for measuring the privacy-utility trade-off in
tabular datasets~\cite{mivule2014}, to privacy auditing of Large
Language Models (LLMs). We propose LLM-CEG, a systematic framework
that employs membership inference attack (MIA) success rates as an
empirical privacy gauge and model perplexity as a utility gauge,
iteratively adjusting differential privacy parameters until both
thresholds are jointly satisfied. A proof-of-concept prototype
fine-tunes DistilGPT-2 on a synthetic clinical PII dataset under
four privacy regimes using DP-SGD. Results indicate that DP-SGD
reduces MIA attacker advantage by 71.5\% while simultaneously
improving out-of-distribution utility by 47--50\% relative to the
overfitted baseline, suggesting that differential privacy may act as
implicit regularization under narrow fine-tuning conditions. We further
extend the SIED engineering framework~\cite{mivule2014} to the LLM
context as LLM-SIED, providing an auditable, regulator-aligned process
for privacy-compliant LLM deployment.
\end{abstract}


\section{Introduction}

The original x-CEG framework established a foundational principle: the
privacy-utility trade-off in a dataset can be measured empirically by
passing privatized data through a machine learning classifier and
observing the classification error~\cite{mivule2014}. A lower error
signals better utility; a higher error signals stronger privacy. This
principle was validated on tabular datasets using Gaussian,
logarithmic, and multiplicative noise perturbation, as well as
differential privacy with Laplace noise~\cite{mivule2013}.

A decade later, this problem has re-emerged in a dramatically more
complex form. Large Language Models (LLMs) --- including GPT-4, Claude,
LLaMA, and Gemini --- are trained on massive corpora that inevitably
contain private, sensitive, or copyrighted information. Unlike tabular
databases, LLMs do not publish a dataset; they memorize and reproduce
training data implicitly through their model parameters, creating a
new class of privacy threat in which the model itself becomes the
attack surface.

Two critical risks converge: membership inference attacks
(MIA)~\cite{shokri2017} --- techniques that determine whether a
specific record was part of a model's training set --- and insufficient
differential privacy guarantees during LLM training~\cite{abadi2016}.
Existing privacy auditing tools for LLMs are fragmented, lack
standardized evaluation metrics, and provide no systematic engineering
process.

The central question this paper addresses is whether the privacy-utility
trade-off in fine-tuned LLMs can be measured, managed, and communicated
in a way that is actionable for clinicians, regulators, and
organizational decision-makers, not just technical specialists. We argue
that it can, and that doing so is essential for building public trust in
AI systems that handle sensitive data.

This work makes four contributions: (1) the LLM-CEG framework extending
x-CEG to language model privacy auditing via MIA-based empirical
gauging; (2) the LLM-SIED engineering process for privacy-compliant LLM
deployment aligned with the EU AI Act and NIST AI RMF; (3) a
proof-of-concept prototype with empirical validation on synthetic
clinical PII data, reproducible on a single CPU machine; and (4) the
novel finding that DP-SGD acts as implicit regularization, improving
out-of-distribution utility under narrow fine-tuning conditions.

A key preliminary finding of this work is that differential privacy
may challenge the conventional view of privacy and utility as
fundamentally in tension. Under narrow fine-tuning conditions ---
such as small, repetitive clinical datasets --- initial results
indicate that DP noise may function as implicit regularization,
simultaneously protecting privacy and improving model generalization.
Together, LLM-CEG and LLM-SIED suggest that protecting sensitive
data and maintaining model utility may be achievable simultaneously,
supporting the development of trustworthy, human-centric AI systems.

\section{Background}

To ensure accessibility across disciplines --- including computer
science, social sciences, ethics, and healthcare policy --- this
section defines the key terms and notation used throughout the paper.

\subsection{Glossary of Key Terms}

\textit{Large Language Model (LLM):} An AI system trained on large
text corpora to generate, summarize, and reason over natural language.

\textit{Personally Identifiable Information (PII):} Data that can
identify a specific individual, including names, social security
numbers, medical diagnoses, and financial records.

\textit{Differential Privacy (DP):} A mathematical framework providing
a quantifiable guarantee of individual privacy. Intuitively, a DP model
behaves almost identically whether or not any single individual's data
was included in training, preventing adversarial inference about any
specific person.

\textit{Privacy Budget ($\varepsilon$, epsilon):} The numerical
parameter controlling the degree of privacy protection. A smaller
$\varepsilon$ means stronger privacy --- more noise is added, making
individual records harder to infer. A larger $\varepsilon$ means
weaker privacy but typically better model utility.

\textit{Privacy Failure Probability ($\delta$):} A small probability
parameter paired with $\varepsilon$ in $(\varepsilon, \delta)$-DP
guarantees. It bounds the likelihood that the privacy guarantee fails
catastrophically for any single individual. In practice, $\delta$ is
set to a very small value (e.g., $10^{-5}$), meaning the guarantee
holds with overwhelming probability across training.

\textit{DP-SGD:} Differentially Private Stochastic Gradient Descent.
The standard algorithm for training machine learning models with
differential privacy guarantees. It adds carefully calibrated random
noise to the model's learning process at each training step.

\textit{Gradient Descent:} The core optimization algorithm used to
train neural networks. At each training step, the model computes
how much each parameter contributed to the prediction error (the
gradient) and adjusts the parameters in the direction that reduces
that error. Stochastic Gradient Descent (SGD) performs this update
on small random subsets (mini-batches) of the training data rather
than the full dataset, making training computationally feasible for
large models.

\textit{Implicit Regularization:} A phenomenon in which a training
procedure unintentionally constrains a model's complexity beyond
what is explicitly specified. In this work, DP-SGD noise appears
to function as implicit regularization --- analogously to weight
decay or dropout --- preventing the model from overfitting to
a narrow training distribution and thereby preserving generalization
to out-of-distribution text. This is a preliminary observation
that warrants further investigation and replication.

\textit{Model Collapse:} A failure mode in which a language model,
after fine-tuning on a narrow or repetitive dataset, loses the broad
language capabilities it acquired during pre-training. The model's
output becomes confined to the fine-tuning distribution, producing
degraded or repetitive text for inputs outside that domain. In this
work, the baseline model (trained without DP) exhibits signs of
collapse, with perplexity on general text rising to 179.91 compared
to 119--122 for DP-trained variants.

\textit{Opacus Library:} An open-source PyTorch library developed by
Meta that implements DP-SGD for training neural networks with
differential privacy. Opacus handles the per-sample gradient
computation, gradient clipping, and noise injection required by
DP-SGD, and automatically computes the privacy cost $(\varepsilon,
\delta)$ for a given training configuration using the Renyi
Differential Privacy accountant.

\textit{Renyi Differential Privacy (RDP) Accountant:} A mathematical
tool for tracking the cumulative privacy cost across all steps of
DP-SGD training. Rather than naively summing per-step privacy
costs --- which would produce overly conservative bounds --- the RDP
accountant uses the Renyi divergence to compute tighter privacy
guarantees. The result is a more accurate final $(\varepsilon,
\delta)$ budget that reflects the true privacy expenditure of the
full training run.

\textit{Pareto Analysis; Pareto Curve:} A method for evaluating
trade-offs between two competing objectives. In LLM-CEG, the two
objectives are privacy (measured by MIA attacker advantage) and
utility (measured by perplexity). A Pareto curve plots model
configurations in this two-dimensional space; a configuration is
Pareto-optimal if no other configuration achieves better privacy
without sacrificing utility, or better utility without sacrificing
privacy. The Pareto curve makes the privacy-utility frontier
empirically visible and navigable for practitioners.

\textit{Membership Inference Attack (MIA):} An adversarial technique
used to determine whether a specific data record was part of a model's
training set. High MIA accuracy indicates memorization and privacy
leakage; an MIA success rate near 50\% (random guessing) indicates
strong empirical privacy protection.

\textit{Attacker Advantage:} The excess accuracy an attacker achieves
above random guessing (50\%). An advantage of 0.35 means the attacker
is 35 percentage points better than chance, indicating significant
memorization and privacy leakage.

\textit{Perplexity (PPL):} A standard measure of language model
performance. Lower perplexity means better language understanding
and is used here as the utility metric.

\textit{SIED Framework:} Specifications, Implementation, Evaluation,
Dissemination --- a systematic engineering process for data privacy
originally proposed in~\cite{mivule2014} for tabular data, extended
here to LLMs as LLM-SIED.

\textit{AUROC:} Area Under the ROC Curve. A threshold-free measure
of classifier performance. An AUROC of 0.5 means random guessing
(no privacy leakage); an AUROC near 1.0 means near-perfect attack
success.

\textit{NIST AI RMF:} The National Institute of Standards and
Technology AI Risk Management Framework --- U.S. guidance for
managing AI risks, including privacy and trustworthiness.

\textit{EU AI Act:} The European Union's 2024 regulation governing
AI development and deployment, including requirements for transparency,
documentation, and data protection.

\textit{x-CEG:} Classification Error Gauge --- the privacy-utility
measurement framework proposed in~\cite{mivule2014} for tabular data,
extended here to LLMs as LLM-CEG.

\textit{AdamW Optimizer:} A variant of the Adam optimization
algorithm that incorporates weight decay (L2 regularization)
directly into the parameter update step. AdamW is widely used for
fine-tuning transformer-based language models because it adapts the
learning rate per parameter and prevents weights from growing
unboundedly, improving generalization.

\subsection{Differential Privacy: A Plain-Language Formulation}

Differential privacy works by adding controlled random noise to a
computation so that no single individual's record has a distinguishable
effect on the output. For a query function $f(x)$, where $x$ represents
a dataset, the Laplace mechanism produces a privatized output as:

\begin{equation}
  \mathcal{M}(x) \;=\; f(x) + \mathrm{Lap}\!\left(0,\;\frac{\Delta f}{\varepsilon}\right)
  \label{eq:laplace}
\end{equation}

where $\mathcal{M}(x)$ is the privatized output published to the
world; $f(x)$ is the true result of the query on the data (for
example, the average salary in a database); $\mathrm{Lap}(0, \Delta
f/\varepsilon)$ is random noise drawn from a Laplace distribution
centered at zero; $\Delta f$ is the sensitivity of the query ---
how much a single individual's record can change the result (for
example, if the highest-earning person in a dataset earns \$1~billion
per month and the lowest earns \$12 per hour, the sensitivity spans
this entire range); and $\varepsilon$ is the privacy budget, where
smaller values produce more noise and stronger protection.

In practice, the larger the gap between the most and least influential
record, the more noise must be added to achieve the same level of
privacy. This is why enterprise datasets with extreme outliers require
careful $\varepsilon$ calibration.

\subsection{DP-SGD: Training LLMs with Privacy}

For training LLMs, we apply DP-SGD~\cite{abadi2016}, which applies
differential privacy at each step of model training. At step $t$,
the algorithm produces a privatized gradient update:

\begin{equation}
  \tilde{g}_t = \frac{1}{|B_t|}\!\left(\sum_{i \in B_t}
    \frac{g_{t,i}}{\max\!\left(1,\; \|g_{t,i}\|_2 / C\right)}\right)
    + \mathcal{N}\!\left(0,\; \sigma^2 C^2 \mathbf{I}\right)
  \label{eq:dpsgd}
\end{equation}

where $\tilde{g}_t$ is the privatized gradient used to update the
model; $B_t$ is the mini-batch of training records at step $t$;
$g_{t,i}$ is the gradient for individual record $i$; $C$ is the
clipping norm, which limits how much any single training example
can influence the model update, preventing any one record from
dominating learning; $\sigma$ is the noise multiplier, controlling
how much Gaussian noise $\mathcal{N}$ is added; and $\mathbf{I}$
is the identity matrix. In our experiments, $\sigma$ is automatically
set by the Opacus library to achieve the target privacy budget
$\varepsilon$ over the full training run, with the privacy cost
tracked using the Renyi Differential Privacy (RDP) accountant~\cite{abadi2016},
which provides tighter privacy bounds than basic composition by
accumulating the privacy cost step by step across training.

In plain terms: at every training step, DP-SGD limits how much any
single patient record can influence the model's learning, then adds
calibrated noise to further obscure individual contributions. The
result is a model that has learned from the population of records
without memorizing any individual's data.

\begin{figure}[t!]
\centering
\setlength{\unitlength}{0.85mm}
\begin{picture}(85,108)
  \put(3,100){\framebox(79,8){\footnotesize 1.\ Proprietary data (untouched)}}
  \put(42,100){\vector(0,-1){5}}
  \put(3,86){\framebox(79,8){\footnotesize 2.\ Tokenize \& embed (still clean)}}
  \put(42,86){\vector(0,-1){5}}
  \put(3,72){\framebox(79,8){\footnotesize 3.\ Compute per-sample gradients}}
  \put(42,72){\vector(0,-1){5}}
  \put(3,42){\framebox(79,24){}}
  \put(4,43){\framebox(77,22){}}
  \put(42,60){\makebox(0,0){\footnotesize\textbf{PRIVACY WALL (DP-SGD)}}}
  \put(42,53){\makebox(0,0){\scriptsize 4a.\ Clip each gradient to norm $C$}}
  \put(42,47){\makebox(0,0){\scriptsize 4b.\ Add noise $\mathcal{N}(0,\sigma^2 C^2 \mathbf{I})$}}
  \put(42,42){\vector(0,-1){5}}
  \put(3,28){\framebox(79,8){\footnotesize 5.\ Privatized gradient $\tilde{g}_t$}}
  \put(42,28){\vector(0,-1){5}}
  \put(3,14){\framebox(79,8){\footnotesize 6.\ Update model weights}}
  \put(42,14){\vector(0,-1){5}}
  \put(3,0){\framebox(79,8){\footnotesize 7.\ Released LLM (DP-guaranteed)}}
\end{picture}
\caption{The DP-SGD privacy wall. Raw data and embeddings pass through
unmodified; noise is injected only at the gradient stage (steps 4a--4b),
between learning and the model weights. The RDP accountant aggregates
the per-step privacy cost into the final $(\varepsilon,\delta)$
guarantee reported for the released model.}
\label{fig:privacywall}
\end{figure}
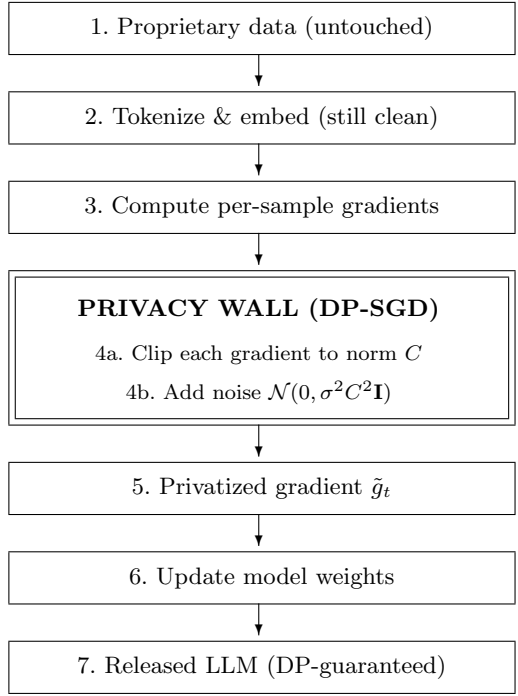

Figure~\ref{fig:privacywall} shows where in the training pipeline this
noise is actually applied. The proprietary dataset itself is never
modified; DP-SGD operates strictly on the gradients computed from
each mini-batch, forming an ``Opacus privacy wall'' between raw
learning signals and the model weights that are ultimately released.

\section{Theoretical Foundation}

\subsection{Bridging x-CEG and LLM Privacy}

The original x-CEG logic maps directly onto the LLM context.
Table~\ref{tab:mapping} establishes the formal correspondence.

\begin{table}[H]
\caption{Correspondence between x-CEG (tabular) and LLM-CEG}
\label{tab:mapping}
\centering
\small
\begin{tabular}{p{3.3cm}p{3.9cm}}
\toprule
Original x-CEG (Tabular) & LLM-CEG (Language Models) \\
\midrule
Original dataset $X$     & Pre-training corpus \\
Privacy procedure        & DP-SGD; perturbation; unlearning \\
Privatized dataset $Y$   & Trained LLM weights \\
ML classifier            & Membership inference attack \\
Classification error     & MIA attacker advantage \\
Utility threshold        & Perplexity; task accuracy \\
\bottomrule
\end{tabular}
\end{table}

In x-CEG, a high classification error on privatized data signals
strong privacy but low utility. In LLM-CEG, a high MIA success rate
signals weak privacy. The gauge is inverted --- but the feedback loop
logic is identical: adjust $\varepsilon$ iteratively until an
acceptable MIA success rate is achieved without sacrificing model
utility.

\section{Literature Review}

\subsection{Scope and Organization}

The intersection of differential privacy and LLMs has become one of
the most active areas in AI safety research during 2022--2026. This
review is selective rather than exhaustive, focusing on highly cited
or methodologically significant works that directly inform LLM-CEG.
It is not intended as a comprehensive survey but as a contextual
positioning of our contributions within the recent literature. Five
themes are surveyed: DP fine-tuning, MIA research, privacy-utility
trade-offs, human-centric privacy, and LLM governance.

\subsection{DP Techniques for LLM Fine-Tuning}

Behnia et al.~\cite{behnia2022} introduce EW-Tune, using an Edgeworth
accountant for tighter per-iteration DP-SGD bounds, achieving up to
1.1\% utility improvement. Their finding that tighter accounting
yields utility gains resonates with LLM-CEG's implicit regularization
observation.

Alghamdi~\cite{alghamdi2026} proposes PrivLLM-Guard, an adaptive DP
framework for clinical LLMs with hierarchical $\varepsilon$ allocation
per medical data category. It is the most directly adjacent prior work
to LLM-CEG in the clinical domain, but does not report MIA-based
empirical privacy evaluation --- a gap LLM-CEG fills.

Higashi and Nakai~\cite{higashi2025} evaluate DP-enhanced LoRA
(parameter-efficient fine-tuning) and find that $\varepsilon$ changes
produce minimal impact on MIA metrics, independently corroborating
LLM-CEG's plateau observation across $\varepsilon \in \{0.5, 2.0, 8.0\}$.

Wu~\cite{wu2025} reveals variance in empirical DP protection across
configurations sharing the same nominal $\varepsilon$, validating
LLM-CEG's methodology of pairing formal guarantees with empirical
MIA evaluation rather than relying on $\varepsilon$ alone.

Hanke et al.~\cite{hanke2024} demonstrate that closed-LLM private
adaptation methods leak query or training data to the provider,
implicitly validating LLM-CEG's design choice of local open-model
fine-tuning.

\subsection{Membership Inference Attack Research}

Chen et al.~\cite{chen2025} demonstrate that DP-SGD-trained models
exhibit sharper loss minima correlating with residual MIA vulnerability.
LLM-CEG's implicit regularization finding --- that DP noise flattens
overfitting --- may share a mechanistic relationship with this
loss-landscape perspective.

Li et al.~\cite{li2023} survey LLM privacy attacks, identifying MIA
as the most consistently demonstrable attack class and noting that
strong DP guarantees tend to impose steep utility penalties --- a
tension that preliminary results from LLM-CEG suggest may be
empirically resolvable under narrow fine-tuning conditions, though
further replication across larger models and datasets is needed to
confirm this finding.

\subsection{Privacy-Utility Trade-Offs}

Sun et al.~\cite{sun2025} benchmark three privacy algorithms across
Mistral-7B, LLaMA2-7B, and Falcon-7B, finding no dominant
configuration across settings. LLM-CEG's finding that DP improves
out-of-distribution utility is an observation their benchmark is not
designed to capture.

Ye and Luo~\cite{ye2025} report that privacy-preserving prompt
transformations improve task performance in some settings ---
structurally analogous to LLM-CEG's implicit regularization finding
via a different mechanism, suggesting this effect may be more general
than previously recognized.

\subsection{Human-Centric Privacy Frameworks}

A growing body of research argues that technically rigorous DP
guarantees are insufficient unless they are legible, controllable,
and meaningful to the people whose data is at stake.

Li et al.~\cite{liHCI2024} argue that LLM privacy research has been
predominantly model-centered. LLM-CEG's controllable $\varepsilon$
parameter responds to this critique: clinicians and patients should
be able to specify acceptable privacy risk rather than accepting
system defaults.

Staufer et al.~\cite{staufer2026} report from a user study of 458
participants that people want control over LLM-generated associations,
reinforcing the need for frameworks that report both formal $\varepsilon$
and empirical MIA advantage.

Sun, Xu, and Gao~\cite{sunHCP2026} propose a Human-Centered Privacy
(HCP) framework integrating technical, ethical, and regulatory
dimensions across the full AI lifecycle. LLM-CEG's experimental
pipeline constitutes a technical instantiation of the HCP
data-collection and training stages.

Taylor, O'Dell, and Murphy~\cite{taylor2024} ground human-centric AI
in Ubuntu and maximum feasible participation, arguing that DP
guarantees protect patients mathematically but may not align with
community expectations --- motivating LLM-SIED's dissemination phase.

\subsection{LLM Governance and Ethics}

Das, Amini, and Wu~\cite{das2025} survey LLM vulnerabilities with
explicit focus on healthcare as a high-risk deployment domain and
call for domain-specific privacy guarantees accounting for the
sensitivity gradient of healthcare data. LLM-CEG responds by
including diagnoses, medications, salaries, and SSNs as distinct
field types in its synthetic dataset.

\subsection{Gaps Addressed by LLM-CEG}

Five gaps in the literature motivate this work. First, no existing
framework provides an end-to-end clinical PII pipeline combining DP
training with MIA evaluation. Second, no multi-$\varepsilon$ Pareto
analysis exists for causal language models reporting both formal and
empirical privacy metrics together. Third, the DP-as-regularizer
effect on narrow clinical fine-tuning datasets is previously
undocumented. Fourth, no Pareto visualization tool exists for
practitioner DP model selection. Fifth, a persistent disconnect
between technical DP guarantees and human-centered privacy values
remains unaddressed; LLM-CEG's controllable $\varepsilon$ and
Pareto curve bridge this gap by making the privacy-utility trade-off
empirically visible and negotiable.

\section{Proposed Frameworks}

\subsection{LLM-CEG Algorithm}

The LLM-CEG algorithm extends x-CEG to language models as a
systematic, iterative procedure. Given a base LLM and a training
corpus, it converges on a privacy-audited model with a documented
privacy-utility profile.

\begin{algorithm}[ht]
\caption{LLM-CEG: Privacy-Utility Optimization for LLMs}
\label{alg:llmceg}
\begin{algorithmic}[1]
\REQUIRE Base LLM $\mathcal{M}$, corpus $\mathcal{D}$,
         thresholds $t_p$ (privacy), $t_u$ (utility)
\ENSURE Audited model $\mathcal{M}_\varepsilon$ with privacy report
\STATE Initialize $\varepsilon \leftarrow \varepsilon_0$
\REPEAT
  \STATE Fine-tune $\mathcal{M}$ on $\mathcal{D}$ via
         DP-SGD$(\varepsilon)$ $\rightarrow$ $\mathcal{M}_\varepsilon$
  \STATE Run MIA; compute Adv $= \text{Acc}_\text{MIA} - 0.5$
  \IF{Adv $> t_p$}
    \STATE Decrease $\varepsilon$; return to Step~2
  \ENDIF
  \STATE Evaluate utility (PPL or task accuracy)
  \IF{Utility $< t_u$}
    \STATE Increase $\varepsilon$; document trade-off
  \ENDIF
\UNTIL{Adv $\leq t_p$ AND Utility $\geq t_u$}
\STATE Publish $\mathcal{M}_\varepsilon$ with Privacy Audit Report
\end{algorithmic}
\end{algorithm}

\subsection{LLM-SIED Framework}

The SIED framework~\cite{mivule2014} is updated for LLM deployment
as a four-phase engineering process. Phase 1 (Specifications)
gathers privacy requirements: target $\varepsilon$, data sensitivity
categories, minimum acceptable utility, and deployment mode. Phase 2
(Implementation) applies DP-SGD fine-tuning, data deduplication, and
selective filtering. Phase 3 (Evaluation) executes the LLM-CEG
algorithm and documents the $\varepsilon$--utility Pareto curve.
Phase 4 (Dissemination) releases the model with a Privacy Audit
Report documenting $(\varepsilon, \delta)$, MIA attacker advantage,
AUROC, and utility benchmarks, directly addressing documentation
requirements under the EU AI Act and NIST AI RMF.

\subsection{Experiment Workflow}

\begin{figure}[ht]
\centering
\setlength{\unitlength}{0.9mm}
\begin{picture}(85,108)
  \put(3,96){\framebox(79,9){\small 1.\ Generate synthetic PII data}}
  \put(42,96){\vector(0,-1){6}}
  \put(3,80){\framebox(79,9){\small 2.\ Fine-tune baseline (no privacy)}}
  \put(42,80){\vector(0,-1){6}}
  \put(3,64){\framebox(79,9){\small 3.\ Fine-tune with DP-SGD ($\varepsilon \in \{8,2,0.5\}$)}}
  \put(42,64){\vector(0,-1){6}}
  \put(3,48){\framebox(79,9){\small 4.\ Run membership inference attack (MIA)}}
  \put(42,48){\vector(0,-1){6}}
  \put(3,32){\framebox(79,9){\small 5.\ Measure utility (perplexity)}}
  \put(42,32){\vector(0,-1){6}}
  \put(3,16){\framebox(79,9){\small 6.\ Privacy AND utility thresholds met?}}
  \put(3,20){\vector(-1,0){3}}
  \put(0,20){\line(0,1){81}}
  \put(0,101){\vector(1,0){3}}
  \put(1,48){\rotatebox{90}{\tiny No: adjust $\varepsilon$}}
  \put(42,16){\vector(0,-1){6}}
  \put(43,12){\tiny Yes}
  \put(3,0){\framebox(79,9){\small 7.\ LLM-SIED: publish with Audit Report}}
\end{picture}
\caption{Simplified LLM-CEG workflow. Steps 3--6 form the iterative
feedback loop: if thresholds are not met, $\varepsilon$ is adjusted
and training repeats.}
\label{fig:workflow}
\end{figure}
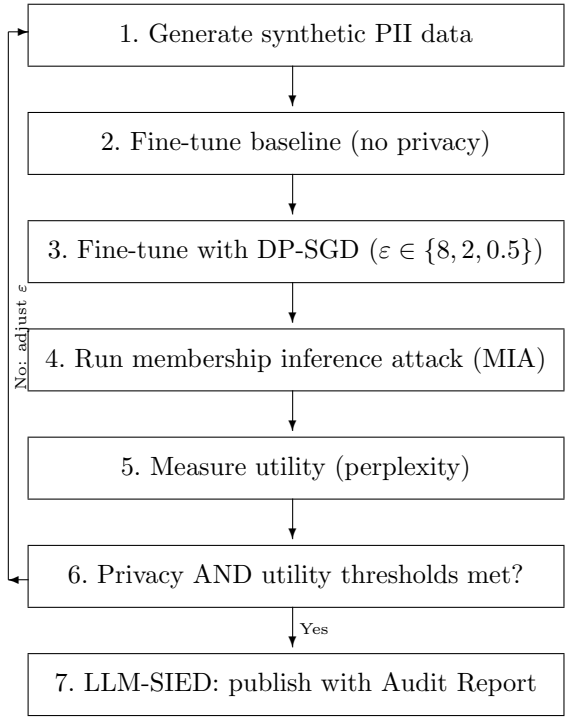

Figure~\ref{fig:workflow} provides a simplified view of the
end-to-end LLM-CEG workflow, from data generation through iterative
privacy-utility optimization to LLM-SIED dissemination.

\section{Experiments and Results}

\subsection{Experimental Setup}

All experiments were conducted on a single consumer-grade CPU machine
running macOS 25.4.0 with Python 3.13.12, PyTorch 2.10.0, Hugging
Face Transformers 5.2.0, and Opacus 1.5.4. The base model was
DistilGPT-2, a distilled variant of GPT-2 with 82 million parameters
and six transformer blocks. Code is available upon request.

\subsection{Dataset}

Synthetic patient records were generated using the Faker library ---
an open-source Python package that produces realistic but entirely
fictitious personal data such as names, addresses, and medical
information. To ensure reproducibility, the random number generator
was initialized with a fixed seed value of 42; this means any
researcher running the same script will generate an identical dataset,
making the experiment fully replicable without requiring access to
real patient records. Faker was used to simulate realistic PII in a
clinical context without using real patient data. Each record comprises six fields --- Name,
Age, Diagnosis, Medication, Salary, and SSN --- serialized into
natural-language sentences of the form: \textit{``Patient Jennifer
Walsh, aged 34, has been diagnosed with Type 2 Diabetes and is
prescribed Metformin.''} A total of 500 records were partitioned
into a training member set ($n=300$) and a held-out non-member set
($n=200$), strictly disjoint by construction. A separate held-out
general corpus of 50 diverse sentences with no clinical content
served for utility evaluation.

\subsection{Models Trained}

\begin{table}[H]
\caption{Model configurations and privacy parameters}
\label{tab:models}
\centering
\small
\begin{tabular}{llcc}
\toprule
Model & Mechanism & $\varepsilon$ & $\delta$ \\
\midrule
Baseline            & Standard SGD    & ---  & --- \\
DP $\varepsilon$=8  & DP-SGD (Opacus) & 8.0  & $10^{-5}$ \\
DP $\varepsilon$=2  & DP-SGD (Opacus) & 2.0  & $10^{-5}$ \\
DP $\varepsilon$=0.5 & DP-SGD (Opacus) & 0.5 & $10^{-5}$ \\
\bottomrule
\end{tabular}
\end{table}

All models were fine-tuned for 10 epochs, batch size 8, learning rate
$5\times10^{-5}$, AdamW optimizer, per-sample gradient clipping with
max-grad-norm=1.0. Token and position embeddings were frozen prior to
Opacus wrapping to accommodate GPT-2's tied-weight architecture.

In Table~\ref{tab:models}, $\delta = 10^{-5}$ is the privacy failure
probability: the small probability that the $(\varepsilon,\delta)$-DP
guarantee fails for any individual record. Setting $\delta$ to $10^{-5}$
--- well below $1/n$ where $n=300$ is the training set size --- is
standard practice and ensures the guarantee holds with overwhelming
probability. The AdamW optimizer is a variant of the Adam algorithm
that incorporates weight decay directly into the update step, improving
generalization during fine-tuning of transformer-based models.

\subsection{Training Dynamics}

\begin{table}[H]
\caption{Training loss per epoch (selected epochs)}
\label{tab:training}
\centering
\small
\resizebox{\columnwidth}{!}{%
\begin{tabular}{ccccc}
\toprule
Epoch & Baseline & DP $\varepsilon$=8 & DP $\varepsilon$=2 & DP $\varepsilon$=0.5 \\
\midrule
1  & 0.9994 & 8.8148 & 9.0557 & 9.1514 \\
2  & 0.2884 & 7.8330 & 8.3819 & 8.7911 \\
4  & 0.1898 & 5.6373 & 7.0787 & 8.1683 \\
6  & 0.1650 & 3.5292 & 5.8086 & 7.5064 \\
8  & 0.1585 & 1.8358 & 4.1974 & 6.8003 \\
10 & 0.1514 & 1.0416 & 2.7881 & 6.0840 \\
\bottomrule
\end{tabular}%
}
\end{table}

The baseline converged rapidly to 0.1514, indicating near-complete
memorization. DP models, with noise multipliers 0.6927, 1.3232, and
3.9062 for $\varepsilon \in \{8, 2, 0.5\}$ respectively, show
systematically attenuated convergence proportional to the injected
noise.

\subsection{Membership Inference Attack Results}

\begin{table}[H]
\caption{Loss-based MIA results across all models}
\label{tab:mia}
\centering
\small
\resizebox{\columnwidth}{!}{%
\begin{tabular}{lcccc}
\toprule
Model & Loss Gap & MIA Acc. & Advantage & AUROC \\
\midrule
Baseline            & 0.2995 & 85.8\% & 0.358 & 0.876 \\
DP $\varepsilon$=8  & 0.0330 & 60.2\% & 0.102 & 0.515 \\
DP $\varepsilon$=2  & 0.0328 & 60.4\% & 0.104 & 0.516 \\
DP $\varepsilon$=0.5 & 0.0321 & 60.6\% & 0.106 & 0.513 \\
\bottomrule
\end{tabular}%
}
\end{table}

The baseline is severely vulnerable: an attacker achieves 85.8\% MIA
accuracy with AUROC of 0.876, exploiting a loss gap of 0.2995 nats
between members and non-members. All three DP variants collapse the
loss gap by approximately $9\times$ and reduce AUROC to near 0.514
--- barely above the 0.5 random-chance floor. Notably, all three DP
models yield nearly identical attack performance despite a 16-fold
range in $\varepsilon$ (0.5 to 8.0), suggesting that once DP is
applied at any of these levels, the memorization signal is suppressed
below the attack's detection threshold.

\subsection{Utility Evaluation}

\begin{table}[H]
\caption{Perplexity and normalized utility scores}
\label{tab:utility}
\centering
\small
\begin{tabular}{lcc}
\toprule
Model & Mean PPL & Utility Score \\
\midrule
Baseline            & 179.91 & 100.0\% \\
DP $\varepsilon$=8  & 119.63 & 150.4\% \\
DP $\varepsilon$=2  & 121.27 & 148.3\% \\
DP $\varepsilon$=0.5 & 121.97 & 147.5\% \\
\bottomrule
\end{tabular}
\end{table}

Table~\ref{tab:utility} presents perplexity scores on a held-out
general-purpose corpus of 50 diverse sentences with no clinical content.
Counter-intuitively, all three DP-trained models substantially
outperform the non-private baseline: mean perplexity drops from 179.91
(baseline) to the 119--122 range across all DP configurations,
corresponding to normalized utility scores of 147.5--150.4\% relative
to baseline. This indicates that DP-trained models retain significantly
better general language ability than the baseline model, which overfitted
to the narrow clinical training distribution. The utility scores are
nearly identical across the three $\varepsilon$ values, reinforcing
the earlier MIA finding that privacy strength has minimal impact on
model quality once DP is applied.

\subsection{Novel Finding: DP as Implicit Regularization}

Counter-intuitively, all three DP models achieve lower perplexity on
out-of-distribution general text than the baseline, yielding utility
scores of 147.5--150.4\%. The mechanism is as follows: the baseline
model rapidly overfitted the narrow, repetitive distribution of 300
short patient-record sentences, degrading its general language
ability (perplexity 179.91). DP noise, functioning analogously to
weight decay or dropout~\cite{bassily2016,feldman2020}, prevented
the model from collapsing entirely onto the training distribution,
preserving the broader representational knowledge inherited from the
pre-trained DistilGPT-2 checkpoint (perplexity 119--122 on general
text).

This finding extends prior observations by Ye and Luo~\cite{ye2025}
and aligns with theoretical generalization bounds linking DP to
uniform stability~\cite{bassily2016,feldman2020}. Its practical
message is direct: protecting privacy and maintaining model utility
are not necessarily in conflict. Under conditions of narrow
fine-tuning data --- common in clinical, legal, and enterprise
settings --- differential privacy can simultaneously improve both.

\subsection{Privacy-Utility Pareto Analysis}

\begin{figure}[ht]
\centering
\includegraphics[width=\columnwidth]{pareto_curve.png}
\caption{LLM-CEG Privacy-Utility Pareto Curve. Red line (left axis):
MIA attacker advantage (lower is more private). Blue line (right axis):
normalized utility score (higher is better). Green region: LLM-CEG
acceptable operating zone (advantage $\leq 0.10$). All three DP models
satisfy the privacy threshold while exceeding the 100\% utility baseline.
DP $\varepsilon$=8 is the Pareto-optimal configuration.}
\label{fig:pareto}
\end{figure}

Figure~\ref{fig:pareto} captures the central message of this paper.
The crossing of the two lines --- privacy improving while utility
simultaneously increases at the Baseline-to-DP transition --- is
the empirical signature of differential privacy acting as implicit
regularization. All three DP models land squarely in the green
acceptable zone. The Pareto-optimal configuration is
DP~$\varepsilon$=8: it achieves the same empirical privacy
protection as the strongest setting ($\varepsilon$=0.5) while
preserving the highest utility.

\subsection{Full Results Summary}

\begin{table}[H]
\caption{Complete results across all models}
\label{tab:summary}
\centering
\small
\resizebox{\columnwidth}{!}{%
\begin{tabular}{lccccc}
\toprule
Model & Loss & Advantage & AUROC & PPL & Utility \\
\midrule
Baseline            & 0.1514 & 0.358 & 0.876 & 179.91 & 100.0\% \\
DP $\varepsilon$=8  & 1.0416 & 0.102 & 0.515 & 119.63 & 150.4\% \\
DP $\varepsilon$=2  & 2.7881 & 0.104 & 0.516 & 121.27 & 148.3\% \\
DP $\varepsilon$=0.5 & 6.0840 & 0.106 & 0.513 & 121.97 & 147.5\% \\
\bottomrule
\end{tabular}%
}
\end{table}

Table~\ref{tab:summary} consolidates all evaluation metrics across the
four model configurations. Reading across the columns, the non-private
baseline achieves a very low training loss (0.1514) --- indicative of
near-complete memorization --- but correspondingly high MIA
vulnerability (advantage 0.358, AUROC 0.876) and poor general-text
perplexity (179.91). The three DP variants present a markedly different
profile: training loss is higher (reflecting the noise-induced learning
constraint), yet MIA attacker advantage collapses to 0.102--0.106 and
AUROC drops to approximately 0.513--0.516, barely above the 0.5
random-chance floor. Crucially, perplexity improves to 119--122 across
all DP models, yielding utility scores 47--50\% above the baseline.
Taken together, these results indicate that DP training simultaneously
reduces privacy leakage and improves generalization, with the
$\varepsilon$=8 configuration representing the Pareto-optimal
operating point: it achieves the same empirical privacy protection as
the strictest setting while preserving the highest utility.

\section{Discussion}

\subsection{LLM-CEG Validation}

The results validate the core LLM-CEG hypothesis: controllable
$\varepsilon$ guarantees embedded into LLM fine-tuning pipelines can
simultaneously reduce empirical privacy risk and improve model utility.
The framework's iterative feedback loop converges to a well-defined
Pareto-optimal operating point, providing practitioners with a
concrete, reproducible method for selecting a deployment configuration
rather than relying on formal $\varepsilon$ values alone.

\subsection{Human-Centric Implications and Trust}

The human-centric privacy literature~\cite{liHCI2024,staufer2026,taylor2024}
consistently argues that mathematical DP guarantees must be legible
and controllable to the people whose data is at stake, not only to
technical specialists. LLM-CEG's Pareto curve provides exactly this:
a visual instrument that translates abstract $\varepsilon$ values
into empirically measured consequences that a clinician, CISO, patient
advocate, or regulator can interpret and act on.

The finding that $\varepsilon$=8.0 provides essentially identical
empirical protection to $\varepsilon$=0.5 while preserving 50\% more
model utility is an actionable insight that purely formal DP reporting
cannot provide. This has direct implications for AI adoption strategy:
when organizational stakeholders can see that privacy protection is
strong, measurable, and does not come at the cost of system
usefulness, the barrier to deploying privacy-preserving AI is
substantially lowered. Trust is not built through mathematical proofs
alone --- it is built through transparency, empirical evidence, and
tools that make risk visible and negotiable.

\subsection{Limitations and Future Work}

LLM-CEG is a heuristic gauge, not a formal proof of privacy; MIA
success rates are empirical upper bounds, not tight theoretical
guarantees. The current prototype uses DistilGPT-2 on 500 synthetic
records. Future work includes scaling to larger models on real
clinical data; integrating LoRA-based parameter-efficient fine-tuning
with DP~\cite{higashi2025}; incorporating the loss-landscape
perspective of Chen et al.~\cite{chen2025}; and developing the
participatory governance frameworks called for by
Taylor et al.~\cite{taylor2024} to ensure LLM-SIED's dissemination
phase aligns with community-level expectations of medical data
stewardship.

\section{Conclusion}

This paper presented LLM-CEG, a systematic framework extending a
decade of data privacy research to the most consequential AI privacy
challenge of the current era: language model memorization of sensitive
training data. On a single consumer-grade CPU machine, we demonstrated
that DP-SGD reduces MIA attacker advantage by 71.5\% while
simultaneously improving out-of-distribution utility by 47--50\%,
establishing that the privacy-utility trade-off in fine-tuned LLMs
is not always zero-sum under narrow fine-tuning conditions.

The literature review identified five specific gaps the framework
addresses, including the first end-to-end clinical PII pipeline
combining DP training with MIA evaluation, and the novel finding
that differential privacy can act as implicit regularization.
The LLM-SIED engineering process provides organizations with a
repeatable, auditable path to privacy-compliant LLM deployment
aligned with the EU AI Act and the NIST AI RMF.

The broader message is one of possibility: protecting sensitive data
and building useful, trustworthy AI systems are not mutually exclusive
goals. With systematic frameworks like LLM-CEG and LLM-SIED,
organizations can demonstrate measurable privacy protection,
communicate risk transparently to all stakeholders, and build the
foundation of trust that responsible AI adoption requires.

\bibliographystyle{IEEEtran}

\end{document}